\documentclass[letterpaper,twocolumn,10pt]{article}
\pdfoutput=1
\usepackage{usenix2019_v3}

\usepackage{adjustbox}
\usepackage{amsmath}
\usepackage{multirow}
\usepackage{graphicx}
\usepackage{pgfplots}
\usepgfplotslibrary{colorbrewer}
\usepgfplotslibrary{groupplots}
\usetikzlibrary{calc} 
\usepackage{siunitx}
\usepackage{xspace}
\usepackage[]{quoting}
\usepackage{arydshln}

\newenvironment{newitemize}{%
\begin{list}{\mbox{}\hspace{5pt}$\bullet$\hfill}{\labelwidth=15pt%
\labelsep=4pt \leftmargin=12pt \topsep=3pt%
\setlength{\listparindent}{\saveparindent}%
\setlength{\parsep}{\saveparskip}%
\setlength{\itemsep}{3pt} }}{\end{list}}

\newlength{\saveparindent}
\newlength{\saveparskip}
\newcommand{\gamesfontsize}{\small}

\newcommand{\hpagess}[4]{
        \begin{tabular}[t]{c@{\hspace*{.5em}}c}
        \adjustbox{valign=c}{\begin{minipage}[t]{#1\textwidth}\gamesfontsize #3 \end{minipage}}
        &
        \adjustbox{valign=c}{\begin{minipage}[t]{#2\textwidth}\gamesfontsize #4 \end{minipage}}
        \end{tabular}
    }

\renewcommand{\paragraph}[1]{\vspace*{6pt}\noindent\textbf{#1}\;}

\relax

\newcommand{\marriagebuilders}{Forum~A\xspace}
\newcommand{\deception}{Forum~B\xspace}
\newcommand{\reddit}{Forum~C\xspace}
\newcommand{\mobile}{Forum~D\xspace}
\newcommand{\spy}{Forum~E\xspace}
\newcommand{\marriagebuildersshort}{A\xspace}
\newcommand{\deceptionshort}{B\xspace}
\newcommand{\redditshort}{C\xspace}
\newcommand{\mobileshort}{D\xspace}
\newcommand{\spyshort}{E\xspace}

\begin{document}

\title{\Large \bf The Tools and Tactics Used in Intimate Partner
Surveillance:\\An Analysis of Online Infidelity Forums}

\author{
    {\rm Emily Tseng$^1$,
    Rosanna Bellini$^2$,
    Nora McDonald$^3$,
    Matan Danos$^4$,}\\
    {\rm Rachel Greenstadt$^5$,
    Damon McCoy$^5$,
    Nicola Dell$^6$ and
    Thomas Ristenpart$^6$
    }\\
    \\
    $^1$Cornell University, $^2$Open Lab, Newcastle University, $^3$University of Maryland, Baltimore County,\\ $^4$Weizmann Institute of Science, $^5$New York University, $^6$Cornell Tech
}

\maketitle

\begin{abstract}

Abusers increasingly use spyware apps, account compromise, and social engineering to surveil their intimate partners, causing substantial harms that can culminate in violence. This form of privacy violation, termed intimate partner surveillance (IPS), is a profoundly challenging problem to address due to the physical access and trust present in the relationship between the target and attacker. While previous research has examined IPS from the perspectives of survivors, we present the first measurement study of online forums in which (potential) attackers discuss IPS strategies and techniques. In domains such as cybercrime, child abuse, and human trafficking, studying the online behaviors of perpetrators has led to better threat intelligence and techniques to combat attacks. We aim to provide similar insights in the context of IPS. We identified five online forums containing discussion of monitoring cellphones and other means of surveilling an intimate partner, including three within the context of investigating relationship infidelity. We perform a mixed-methods analysis of these forums, surfacing the tools and tactics that attackers use to perform surveillance. Via qualitative analysis of forum content, we present a taxonomy of IPS strategies used and recommended by attackers, and synthesize lessons for technologists seeking to curb the spread of IPS.
\end{abstract}

\section{Introduction}

Technology-based intimate partner surveillance (IPS) causes immense harms. A discrete form of intimate partner violence (IPV), IPS is the deliberate surveillance of an intimate partner with or without their knowledge, levied through
technical and non-technical methods. Survivors have reported their abusers use
spyware apps,
account compromise, GPS trackers, shared cellular plans, and more to monitor their digital lives and physical
locations~\cite{freed2017digital,freed2018stalker,matthews2017stories,woodlock2017abuse,southworth2007intimate}. 
Prior work has also indicated that a wealth of IPS apps are
available online~\cite{chatterjee2018spyware} and in active use against
victims~\cite{roundy2020many,havron2019clinical,freed2019clinical}. 

To better protect targets of abuse, we need to both improve technologies'
robustness to abuse and better inform \mbox{interventional} approaches that directly
aid victims~\cite{havron2019clinical,freed2019clinical}. To achieve these aspirations, however, we need to better understand how those interested
in perpetrating IPS learn to conduct these attacks. To date, there have been few investigations into how attackers locate the resources that help them enact abuse. Chatterjee et al.~\cite{chatterjee2018spyware} highlight that blogs, videos, and question-and-answer sites exist online that help facilitate IPS, but stop short of 
investigating the communities who make use of them. There is a
methodological hurdle in discovering this information: we need a way to hear from potential attackers directly. 

In this work, we provide the first study exploring how potential attackers use the Internet to learn how to enact IPS
against their victims. 
We identify a set of five public, online forums where people discuss infidelity in intimate relationships and tools for monitoring cellphones. We build a crawler to
retrieve the conversations on these forums and use it to 
compile a dataset containing over 200\,K posts spread across almost 20\,K threads.
This dataset contains an unprecedented amount of information about the strategies of IPS attackers, contextualized in user-generated natural language. While prior work has described the attacks experienced by victims, we present a detailed view of how these attacks are created and developed---the capabilities attackers seek, the vulnerabilities they exploit, and the community dynamics that enable them.

We analyze this data using mixed-methods. We begin with quantitative measurements of the forums, their users, and their posting behaviors.
These analyses reveal that most forums contain ``superusers'' involved with a
disproportionately large fraction of threads. For two of the five forums, we discover that
their content consists almost entirely of spam advertising particular spyware
tools, in a manner that may be consistent with search engine optimization (SEO) techniques deployed by the creators of those tools. We also
determine that many of their most highly viewed threads are about
technology-based IPS, suggesting these forums have generated a significant audience for IPS-related content. 

We then perform qualitative coding of a large sample of 750 threads (250 from each of the  three forums that are not \mbox{inundated} by spyware ads). 
Via thematic analysis \cite{braun2006using}, we
\mbox{surface} novel insights about the online behaviors of IPS attackers. We show that potential attackers seek online support for suspicions of infidelity, and that community members respond by outlining exactly how to track, monitor, and otherwise compromise the privacy of an intimate partner. We show that discussion of IPS is prevalent in these forums, with one 
forum having 78\% of sampled threads related to IPS.

We develop a taxonomy of IPS attacks surfaced from the suggestions made in these forums (Table~\ref{tab:taxonomy}). \textit{Tool-based attacks} directly weaponize technology such
as audiovisual recorders, keyloggers, backup recovery tools, and more, and can be understood in two subcategories: those requiring physical access to a partner, and those that do not.
\textit{Coercion and subterfuge} attacks manipulate a partner into unlocking their
devices or accounts. Finally, we see many suggestions to \textit{outsource
attacks} by hiring private investigators.

Although
some of these strategies have been reported by victims \cite{freed2018stalker, matthews2017stories}, our analysis
provides the complementary view of potential perpetrators.
We highlight tools, tactics, and services that have not
been reported previously, and which we believe were previously unknown to
those helping victims. We also report on the conversational patterns within these forums that enable would-be attackers (what we describe as \textit{escalation}), and, conversely, patterns in which they are discouraged away from IPS (\textit{de-escalation}). These findings suggest that public forums can serve as a source of threat intelligence informing interventions to dissuade abuse.

In fact, our work is already having impact for IPS interventions. We shared our results with the team running a clinic providing direct assistance to IPV survivors facing IPS \cite{havron2019clinical, freed2019clinical},\footnote{https://www.ipvtechresearch.org} who are working towards using our findings in their procedures. More broadly, our analyses yield insights for technologists, platforms, and advocates on how we might defend against and mitigate the spread of IPS. We close by discussing implications for platforms, future work in automated threat intelligence, and how policymakers might address the for-profit operations
that financially benefit from abuse.

\section{Background and Related Work}

\paragraph{IPV and technology abuse.} Prior work has examined the behaviors, justifications, and tactics of intimate partner abusers \cite{hearn_violences_1998, kelly_naming_2016, stark_coercive_2009}, including work identifying suspicions of infidelity as a leading trigger for IPV in heterosexual couples \cite{arnocky2015anticipated, nemeth2012sexual}. Of this literature, a growing
body of work explores the role of technology in IPV, including how abusers
exploit technology to monitor, harass, control or otherwise harm their
targets~\cite{freed2017digital,freed2018stalker,chatterjee2018spyware,matthews2017stories,dimond2011domestic,woodlock2017abuse,southworth2007intimate}. Chatterjee et al.~\cite{chatterjee2018spyware} observed that abusers are likely exploiting easy-to-find online resources, including
tutorials, question-and-answer sites, and videos explaining how
to use spyware for IPS. Roundy et al.~\cite{roundy2020many} used
datasets from a major antivirus vendor to explore a broader class of creepware
that includes spyware, but also SMS bombers, hacking tutorials, and more. These works have provided valuable intelligence for corresponding anti-IPS interventions with victims and survivors \cite{freed2019clinical, havron2019clinical}.

To date, however, less research has examined the role of online communities in IPV. Some have examined how targets experience IPV in digital media and seek support through online forums \cite{dragiewicz2018technology, leitao2019technology}, but to the best of our knowledge, ours is the first study to measure and analyze how forums lead attackers to such tools. Our work confirms that attackers are discussing and recommending IPS strategies on public forums available to any Internet user. We also identify new tactics, such as custom scripts to monitor websites visited and launch man-in-the-middle attacks.

\paragraph{Online measurement studies.} Prior work has used
measurement and analysis of online forums to shed light on communities
discussing criminal or otherwise malicious behaviors. For some of these communities, this research has led to the development of threat intelligence.

Commercially motivated criminals, such as spammers and black-hat hackers, 
use online forums as marketplaces and for learning adversarial techniques from
each other~\cite{thomas2006underground,motoyama2011analysis,franklin2007inquiry}. Research on this phenomenon has identified structure, trust, and superusers in these communities~\cite{afroz2013honor, afroz2014doppelganger, garg2015fc}. Relatedly, online forums used by pedophiles and others involved in the creation and
distribution of child sexual abuse
materials have been studied to gain insights into the way participants justify
or normalize abuse, and share technical and non-technical strategies that facilitate
abuse~\cite{durkin1999propagandizing}. Similar methods have also
been used to analyze forums associated with 
hate groups~\cite{chatzakou2017hate,ribeiro2018characterizing}, cyberbullying~\cite{kowalski2012cyberbullying,chatzakou2017measuring}, doxxing~\cite{snyder2017fifteen},
mis- and disinformation campaigns~\cite{starbird2019disinformation}, harassment~\cite{mariconti2018you, hua20twitter, hua20towardstwitter}, and sex
trafficking~\cite{portnoff2017backpage}. These measurement studies include ones that directly document abuse
(e.g., hate and harassment on social media), as well as
those investigating perpetrators' discussions of tactics (e.g., perpetrators sharing ways to maximize the emotional impact of harassing messages).

Our work falls in the latter category. While methodologically similar, we
differ in our focus on people who use online forums to discuss strategies for IPS. Similar to the work on cybercrime and child abuse forums, analysis of online resources such as craigslist and backpage has led to threat intelligence that helped combat human trafficking~\cite{portnoff2017backpage}. We aim to have a similar impact on IPS and IPV more broadly.
In summary, our research questions are:
\begin{newitemize}
\item What role do online forums play in surfacing IPS resources to potential attackers?
\item What role do commercially motivated entities play in these online communities? 
\item What tools and tactics are being suggested to potential attackers, and at what levels of technical sophistication?
\end{newitemize}

\section{Forums and Datasets}
\label{sec:methods}

To answer these research questions,
we perform a mixed-methods analysis of a large sample of posts and threads from
public online forums with a high density of IPS-relevant conversation.  In this section, we review our analysis targets and data
collection approach, as well as the resulting datasets. 

\paragraph{Infidelity and IPS forums.} We identified several forums whose
content includes a large number of posts touching on IPS. These were discovered
through a combination of pointers from prior work~\cite{chatterjee2018spyware}
and online web searches using a combination of terms such as ``spyware track wife''.
While we endeavored to be exhaustive, we restricted attention only to publicly
available forums, excluding forums accessible only to registered users or users who had crossed some threshold number of active posts. We may also
have missed forums that are not easily found via search engines. 
Finally, many forums have a small number of posts touching on IPS, but with the overwhelming majority of content being irrelevant to our study. We excluded those from our analysis and selected forums that seemed to have a higher concentration of IPS-related posts. Future work might explore techniques to discover IPS-related forums that are harder to access and find.

Our analyses focus primarily on three forums that aim to help people navigate infidelity. To prevent the publication of this work from advertising spyware, or from unintentionally impacting these public forums by convincing them to go private, we anonymize them.\footnote{Our data, including forum names, is available for research upon request.} The forums we study include:

\begin{newitemize} 
\item \textbf{\marriagebuilders}, a community dedicated to discussing  ``\textit{investigative equipment}.'' \marriagebuilders is a subforum of a website providing resources on resolving marital conflicts with Alexa rank approximately 500,000.
\item \textbf{\deception}, a community dedicated to advice on ``\textit{detecting infidelity and deception.}'' \deception is a subforum of a website providing resources on cheating in romantic relationships, with Alexa rank approximately 900,000.
\item \textbf{\reddit}, a moderated Reddit subforum that bills itself as ``\textit{a safe place to ask for advice and guidance}'' for those facing infidelity in a relationship. As of February 2020, \reddit\ had approximately 80,000 subscribers, and Reddit was the 18th most popular website in the world.\footnote{\url{https://www.alexa.com/siteinfo/reddit.com}. Alexa rankings are based on global traffic and engagement over the last 90 days.}
\end{newitemize}
We additionally investigated two subforums that focus on spyware tools: \textbf{Forums \mobileshort} and \textbf{\spyshort}, both subforums of a community for cellphone advice. \mobile\ focuses on spyware for mobile phones, while \spy\ focuses on spyware generally. These subforums surfaced in Internet searches for the same sets of IPS-related keywords as those used to discover the three infidelity forums above. As we discuss further in Section~\ref{sec:data-desc}, our analysis concludes 
most content on these forums are spam advertisements for particular spyware tools.
For simplicity, we will use the term ``\textit{forum}'' to refer to the communities we studied in-depth (e.g., \reddit), and ``\textit{parent forum}'' to refer to their parents where needed (e.g., Reddit).

\paragraph{Data collection.} We collected data from Forums \marriagebuildersshort, \deceptionshort, \mobileshort and \spyshort via custom crawlers built using Scrapy, an
open-source, Python-based framework for web
scraping.\footnote{\url{https://docs.scrapy.org/en/latest/}} Our crawlers preserved the threaded structure of each forum's content, as well as metadata like views and post and update timestamps where available. We did not download any media beyond text---specifically avoiding images---and stored all data in a local database on a secured server.
Our analysis covers a set of scrapes collected using this pipeline in October of 2019. For \reddit, we used the scrape available via the Reddit corpus within ConvoKit \cite{chang2019convokit}, which was collected in October 2018. 
Table ~\ref{tab:dataset} summarizes the complete dataset.

\paragraph{Limitations.}
Our study combines quantitative and qualitative methodologies to characterize a sampling of publicly available forums where discussion of IPS tactics manifests. We emphasize our work may not generalize to discussion on private forums, such as those that require the creation of an account and a threshold of posts or months active for access, or those occurring within private social groups on larger social media platforms such as Facebook and Twitter. We also focus on English-language forums, and thus our findings cannot be taken to represent the scope of IPS discussion worldwide. 
We believe there is compelling future research in investigating larger public-facing communities, like other subreddits or closed communities on and off of influential social networks.

\paragraph{Ethics.} Throughout this work, we were sensitive to the ethics of
using online discussions of highly personal topics for research. Our data is from publicly available fora accessible on the Internet
without authentication. 
Our IRB office reviewed the study and deemed it to not be human-subjects research, since all data was already in the public domain. Still, we took precautions to ensure our research remained safe and privacy-preserving. The only identifiers we used were the public usernames
associated with each post. We did not pursue identification of
people from their posts or usernames, or collect or store images. In reporting our work, we have scrubbed
information that might trace back to the people behind the pseudonyms, e.g.,
locations or specific narrative details.

\section{Forum Activity and Users}
\label{sec:data-desc}

\begin{table*}[ht]
    \centering
    \footnotesize
    \begin{tabular}{@{}lrrrrr@{}}
    \hline 
     & \multicolumn{1}{c}{\textbf{\marriagebuilders}} &
     \multicolumn{1}{c}{\textbf{\deception}} &
     \multicolumn{1}{c}{\textbf{\reddit}} & \multicolumn{1}{c}{\textbf{\mobile}}
     & \multicolumn{1}{c}{\textbf{\spy}} \\ 
    \hline 
    Date of first thread         & Jan 2006 & Aug 2005 & May 2013 & Oct 2008 & Feb 2013 \\
    Size of forum (threads)      & 268       & 1,175     &  11,291  & 3,388     & 2,788 \\
    Size of forum (posts)        & 1,608     & 8,932     &  183,381 & 7,540     & 4,952 \\
    Unique active users in forum & 462       & 2,102     &  12,740  & 264       & 543 \\
    \hline
    Avg. thread views (stdev)         & 3,438 (13,249)     & 4,822 (12,194)     & -- & 1,685 (7,634)    & 6,315 (44,813) \\
    Avg. thread length in posts (stdev)       & 7 (17) & 4 (8) & 16 (17) & 2 (1) & 2 (2) \\
    Avg. time to new thread (stdev)           & 140 days (198 days)   &  7 days (13 days)   &  3 days (13 days) &  1 day (11 days) & 21 hrs (11 days) \\
    Avg. time to new post (stdev)           & 3 days (13 days)   &  14 hrs (2 days)   & 15 minutes (2 hrs)  & 12 hrs (5 days)  & 12 hrs (2 days) \\
    \hline
    IPS-relevant \% of threads$^\alpha$ & 78 & 51 & 18 & -- & --\\
    Size of IPS-relevant sample (posts)$^\alpha$ & 1,411 & 2,011 & 1,032 & -- & -- \\
    Unique users active in IPS-relevant threads$^\alpha$ & 296 & 465 & 346 & -- & -- \\
    \% of IPS-relevant threads that escalate$^\alpha$ & 32 & 38 & 35 & -- & --\\
    \hline
    \end{tabular}
    \caption{Comparison of the five forums in our dataset. \reddit does not provide viewership
    information (marked by dashes). $^\alpha$\,Calculated
    via qualitative analysis of random samples of 250 threads per non-spam forum, see Section~\ref{sec:qual-methods}.}
    \label{tab:dataset}
\end{table*}

We begin by measuring the nature of activity on these
forums. Later on (Sections~\ref{sec:seeker-experience},~\ref{sec:taxonomy}), we use qualitative methods to more deeply characterize their content.

\paragraph{Forum activity and viewership.} The forums in our dataset varied in their rates of activity and reported viewership. 
In total, the forums
contain 18,937 threads, with \reddit\
containing the most threads and posts (Table~\ref{tab:dataset}). 

\begin{figure*}[t]
    \centering
    \includegraphics{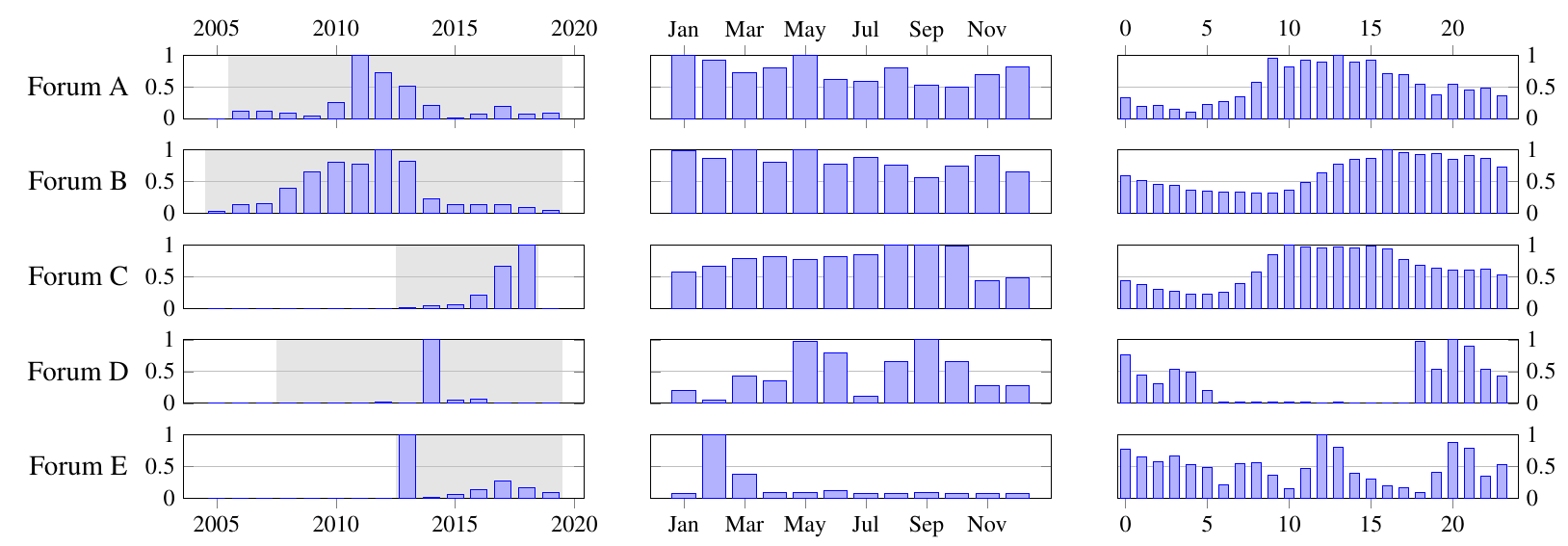}
    \caption{Histograms (normalized to maximum bin value in forum) for 
    \textbf{(left)} postings 
    per year, with shading indicating the years for which we have post data for
    the forum, \textbf{(middle)} postings per month of year, and
    \textbf{(right)} postings per hour of day. }
    \label{fig:forums-over-time}
\end{figure*}

Activity data in terms of thread and post times was available for all forums. We note that activity on Forums \marriagebuildersshort and \deceptionshort peaked between 2010 and 2015, and has significantly dropped off in the last five years, while activity on \reddit has exploded in that time (Figure~\ref{fig:forums-over-time}). We hypothesize this may represent a shift away from niche forums focused on infidelity and towards niche subcommunities of larger social media platforms like Reddit. Despite the recent drop-offs, these forums remain publicly available resources for would-be abusers, and contain IPS tools and tactics that are still relevant today; thus we included them in our qualitative analysis (Section \ref{sec:qual-methods}).

While these three forums exhibit similar seasonal and diurnal patterns, temporal patterns for Forums \spyshort\ and \mobileshort\ exhibit greater variability, as well as strong peaks in year-over-year posting activity in 2013 and 2014, respectively. As we will discuss subsequently, this reflects concentrated activity by advertisers posting spam marketing spyware products.

Across all forums, the total number of views was approximately 30\,M. This is likely a significant underestimate of total viewership given that it does not include Reddit's \reddit, for which we do not have viewership data. Within each forum, the distribution of views per thread was dominated by one or two highly viewed threads (usually `sticky'
threads compiling forum rules or shared resources) and then a long tail of
less-viewed threads. The distributions of thread lengths for each forum followed similar long-tail patterns, with an average thread length of six posts.

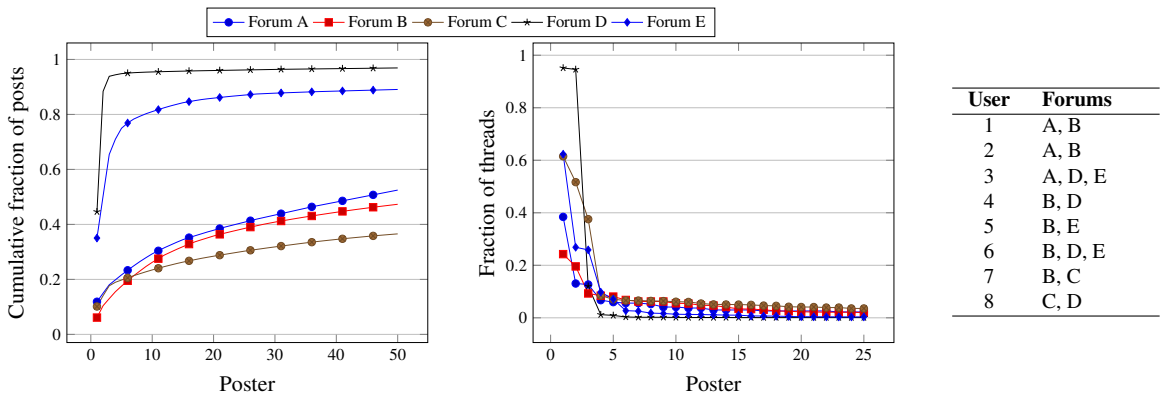
\begin{figure*}[t]
    \centering
    \hpagess{.70}{.25}{
      \pgfplotstableread[col sep = comma]{figures/posts_cdf.csv}\mypostsdata
\pgfplotstableread[col sep = comma]{figures/threads_freq.csv}\mythreaddata
  \begin{tikzpicture}[scale=0.70]
    \begin{axis}[
      name=ax1,
      mark repeat=5,
      legend style={at={(0.5,1.0)},anchor=south},
      ymajorgrids=true,
      xlabel={\large Poster},
      ylabel={\large Cumulative fraction of posts},
      ylabel near ticks,
      legend columns = 5,
      ]
      \addplot table[x={user}, y = {OI}] {\mypostsdata};
      \addlegendentry{\marriagebuilders};

      \addplot table[x={user}, y = {TD}] {\mypostsdata};
      \addlegendentry{\deception};
      
      \addplot table[x={user}, y = {SI}] {\mypostsdata};
      \addlegendentry{\reddit};

      \addplot table[x={user}, y = {MPSS}] {\mypostsdata};
      \addlegendentry{\mobile};

      \addplot table[x={user}, y = {SS}] {\mypostsdata};
      \addlegendentry{\spy};

      \legend{}
    \end{axis}

    \begin{axis}[
      at={(ax1.south east)},
      xshift=2cm,
      legend style={at={(-.2,1.02)},anchor=south},
      ymajorgrids=true,
      xlabel={\large Poster},
      ylabel={\large Fraction of threads},
      ylabel near ticks,
      legend columns = 5,
      ]
      \addplot table[x={user}, y = {OI}] {\mythreaddata};
      \addlegendentry{\marriagebuilders};

      \addplot table[x={user}, y = {TD}] {\mythreaddata};
      \addlegendentry{\deception};
      
      \addplot table[x={user}, y = {SI}] {\mythreaddata};
      \addlegendentry{\reddit};

      \addplot table[x={user}, y = {MPSS}] {\mythreaddata};
      \addlegendentry{\mobile};

      \addplot table[x={user}, y = {SS}] {\mythreaddata};
      \addlegendentry{\spy};

    \end{axis}
\end{tikzpicture}
    }{
      \footnotesize
      \begin{tabular}[c]{clr}
      \hline
      \textbf{User} & \textbf{Forums} \\
      \hline
      1 & \marriagebuildersshort, \deceptionshort & \\
      2 & \marriagebuildersshort, \deceptionshort &\\
      3 & \marriagebuildersshort, \mobileshort, \spyshort&\\
      4 & \deceptionshort, \mobileshort&\\
      5 & \deceptionshort, \spyshort&\\
      6 & \deceptionshort, \mobileshort, \spyshort&\\
      7 & \deceptionshort, \redditshort&\\
      8 & \redditshort, \mobileshort&\\
      \hline
      \end{tabular}
    }
    \caption{\textbf{(Left)} Cumulative fraction of posts per user for the top
    50 users. \textbf{(Middle)} Fraction of threads with posts from the top 25
    users. \spy\ (resp.~\mobile) are dominated by 1 (resp.~2) superusers, while the other three forums show a more even spread of posts and threads among users.
    \textbf{(Right)} Multiple-forum users and the forums in which they posted. Excludes the 17
    users found posting in Forums \mobileshort\ and \spyshort, which share a parent forum.}
    \label{fig:superusers}
    \label{tab:mult-forum-users}
\end{figure*}

\paragraph{Forum users and ``superusers''.} Table~\ref{tab:dataset} shows the number of users in each forum, identified by comparing the usernames attached to each post via case-insensitive string equality. Forums differed in the number of unique users, from 264 in \mobile\ to 12,740 in \reddit, but all forums have ``superusers'' that account for a disproportionate number of
posts. Figure~\ref{fig:superusers} gives (left chart) a CDF of the fraction of each forum's
posts made by users
and (right chart) a histogram of the fractions of all threads to which a user
posted. For clarity, we only show the 50 and 25 most prolific users,
respectively.

Forums \spyshort and \mobileshort are clear outliers compared to the other forums; this is due to spammers, as we discuss below. While the other three forums also have superusers, they do not dominate their forums to the same degree. Additionally, cursory examination shows they are not spammers. Some are human and robot moderators, including an automatic moderator on \reddit that posts the subreddit's rules as the first response to each thread-starting post. But most superusers appear to be humans particularly engaged in the forum, driving the culture and activity of the community with their posts.

We additionally checked whether posters were active in multiple forums in our data. Comparing usernames via case-insensitive string equality, we found just eight users recurring across forums that had no structural reason to be connected. Of these, only one user made contributions that exceeded 1\% of posts or threads in any forum. While this finding seems to indicate superusers are not cross-posting across multiple forums, we note it is simple to register accounts with different usernames in these forums. We consider the identification of users across forums to be an area of future work.

\paragraph{Spyware spam and SEO inflation in Forums \mobileshort\ and \spyshort.} As mentioned above, Forums
\mobileshort and \spyshort stood out along many dimensions. Most content in these two forums can be attributed to
a handful of users: notably, the top user in \mobile\ contributed to 95\%
of threads and authored 45\% of posts, and the second-most-active user
contributed to 95\% of threads and authored 44\% of posts. \spy exhibits a similar pattern of dominance by a handful of users.

Inspection shows many of the threads in \mobile constitute conversations between its top two users: one posts a spam advertisement for a spyware tool, and another follows up with a short response. We conclude this demonstrates a strategy of search engine optimization (SEO) employed by the company behind the spyware tool to boost the forum's visibility on Internet searches and attract attention to their spyware product. Specifically, 94\% of the posts made by the top user were the same message: an advertisement for the spyware tool. This user also authored nearly half (45\%) of the posts on this forum. \mobile's
second-most-prolific user bears a username closely
associated with that spyware tool, and posts either ads for the tool or
short, meaningless messages (\textit{``Hi,''} \textit{``Hello''}) in response to the first user's ads. We found similar patterns within \spy. Having concluded that most activity on these two forums was spam intended to inflate SEO for specific spyware products, we excluded them from our qualitative analysis of forum content (Sections~\ref{sec:seeker-experience},~\ref{sec:taxonomy}). 

\paragraph{Prevalence of IPS-related keywords.} To efficiently understand the organic content in the three infidelity forums, we sought automated ways to identify only those threads that were relevant to IPS. As a
first-cut assessment, we performed keyword-based
searches of the threads in our dataset using a small set of keywords identified from prior work \cite{chatterjee2018spyware}: ``\textit{spy}'',
``\textit{monitor}'', ``\textit{track}'', ``\textit{hack}'' and ``\textit{record}'' (Table~\ref{tab:keyword-relevance}).

This first-cut assessment showed keyword searches are, unsurprisingly,
insufficient for accurate discovery of relevant threads: for example, the keyword `\textit{record}' may be used in the context of recording someone without their consent, but also in the context of music recordings. To quantify this, we assembled a human-labeled dataset of 750 threads sampled across each of the three non-spam forums and manually coded for relevance to IPS (see Section~\ref{sec:qual-methods} for detailed methods). 
We then applied a regex-based labeling method that flagged threads as relevant if any post within the thread contained any one of keywords in our seed set. 
Using our 750 human-labeled threads
as ground truth, this simple approach achieves an AUC of
0.62, indicating it misses a large number of relevant threads (false negatives) and contains a
large number of irrelevant threads (false positives). 

As a result, we do not rely on the regex-based approach for any of our subsequent analyses, but instead study the posts human-labeled as ground truth. The development of automated learning techniques that can efficiently flag IPS-relevant threads remains a tantalizing area of future work.

\begin{table}[t]
\centering
\footnotesize
\begin{tabular}{lrrrrr}
\hline
\textbf{Keywords} & \textbf{\marriagebuildersshort} & \textbf{\deceptionshort} & \textbf{\redditshort} & \textbf{\mobileshort} & \textbf{\spyshort} \\ \hline
spy     & 26.9   & 7.1  & 1.6   & 98.4  & 42.1 \\
monitor & 8.6   & 2.2   & 3.0   & 97.8  & 27.8 \\
track   & 13.4   & 5.4  & 8.8   & 25.7  & 30.0 \\
hack   & 3.7   & 1.1  &  2.3  & 1.1  & 4.1 \\
record   & 14.9   & 7.8  & 7.3   & 3.8  & 1.4 \\
spy, monitor, track, hack, record & 43.7 & 17.3 & 17.6 & 99.4 & 62.3 \\ \hline
\end{tabular}
\caption{Percentage of threads within each forum containing one or more of the indicated keywords.}
\label{tab:keyword-relevance}
\end{table}

\section{Understanding Forum Content}
\label{sec:qual-methods}

Our data contain rich information on attackers' strategies, interactions, and stated goals embedded in the natural language of users' posts. Here, we describe our qualitative methods for analyzing the content within Forums \marriagebuildersshort, \deceptionshort\ \& \redditshort.

\paragraph{Establishing human ratings for IPS relevance.} Our initial measurements showed not all content on these forums is relevant to the discussion of IPS tactics: for example, while 8 of the top 10 threads by viewership on \deception\ contained some mention of ways to monitor an intimate partner, the other two threads discussed contraception and women's underwear. Thus, to focus our analysis, we first established human ratings for whether or not a given thread was relevant to IPS. 

We began by randomly choosing 30 threads, 10 from each forum. Three coders independently rated whether each thread was IPS-relevant.
We stipulated that a relevant thread should both (1) discuss an intent to track, monitor, surveil, or otherwise compromise an intimate partner's privacy; and (2) describe doing so via technology. Inter-rater reliability showed agreement in 28/30 threads (Fleiss' kappa of 0.91 \cite{mchugh2012interrater}). 

We then expanded our analysis to arrive at a set of IPS-relevant threads for further study. We randomly sampled 750 threads (250 from each forum) that we split evenly among the three coders.
As reported in Table~\ref{tab:dataset}, we ultimately found 78\% of the sampled data within \marriagebuilders\ was relevant to IPS; 51\% within \deception; and 18\% within \reddit. These figures are in line with expectations: \marriagebuilders, which is explicitly dedicated to ``\textit{investigative equipment}'', has the highest prevalence of IPS-related content, while \reddit, which has a more general focus on discussion of infidelity, has the lowest.

In total, 370 of the 750 randomly sampled threads were coded as IPS-relevant. We found no statistically significant correlations between thread viewership and IPS relevance in any forum, or any noteworthy patterns in seasonal, diurnal, or year-over-year posting activity within IPS-relevant data.

\paragraph{Understanding IPS-relevant content.} 
We used open thematic coding \cite{braun2006using} to make sense of the 370 IPS-relevant threads.
Three researchers independently read through several threads and generated initial codes. We then met over multiple sessions to jointly develop a codebook. Through multiple rounds of iteration, we refined the codebook by applying it to additional data until we reached saturation and codebook stability.

Our final codebook contained 29 codes clustered into two high-level categories: \textit{forum culture} and \textit{tools and tactics} (see Appendix~\ref{sec:codebook}). Once the codebook was finalized, three researchers divided up the remaining threads and coded them. Our research team stayed in close correspondence throughout the analysis, repeatedly meeting to discuss threads that were unclear at first pass. We also took steps to minimize the impact of repeated readings of detailed stories of IPS and violence on our team. Researchers were encouraged to take breaks where needed, to reach out to each other regularly to process what we were reading, and to practice self-care.

We report the themes that emerged from our analysis in Sections~\ref{sec:seeker-experience} and~\ref{sec:taxonomy}. We emphasize that our analyses are qualitative: thus, we do not report raw or percentage prevalence numbers for any of our themes, except where noted and appropriately tested via inter-rater reliability measurements.

\section{Forum Interactions}
\label{sec:seeker-experience}

In this section, we give a general overview of how users interact within these forums. We begin by discussing how users self-report finding these communities, and what they seek within them. We then describe how communities respond to their requests. We identify threads in which communities encourage users to conduct IPS, either by encouraging them to carry out existing attacks, or by providing them with ideas for attacks of increased severity---what we call \textit{escalation}. We also identify threads in which communities discourage users from IPS at all, a pattern we call \textit{de-escalation}.

\paragraph{How users find these forums.} In several threads in each forum, users describe how and why they sought out these forums in the first place. Many described locating the forum via basic Google queries on topics related to infidelity and cheating. In other posts, users reported discovering the site through a trusted recommendation from a professional enlisted to help them with their relationships, such as a therapist.

Our data also show that for Forums \marriagebuildersshort\ and \deceptionshort, users are often directed to these specific forums by moderators or users of other communities within their parent forum. For example, a moderator in \marriagebuilders posted in response to a thread starter: 

\begin{quoting}[leftmargin = .1 in, vskip= .1 in, indentfirst = false]
    \textit{``I asked you to come here to click on the many threads and read information for yourself. There are pages and pages all about spying \dots This forum is a kind of archive where the information will be available for anyone to peruse at leisure. There is no need to wait. Look around!''}
\end{quoting}

Here, we see a forum moderator reinforce that these discussions are a resource for anyone to browse. We discovered that \marriagebuilders\ in particular hosts several `resource threads' pinned at the top of the forum that provide primers for beginners.

\paragraph{Forum-goers' stated goals.} Once in the forum, most users make an initial post outlining a complex social situation, \mbox{usually} suspected or actualized infidelity, and ask the community for advice on what to do, e.g. ``\textit{How can I forgive him?}'' or ``\textit{How do I move on?}'' Most of these posts sought suggestions from others for `next steps', such as confronting their intimate partner or seeking legal advice. The bulk of posts within \reddit\ followed this pattern, mimicking the advice-seeking observed broadly in forums for social support \cite{yang2019seekers, Fu-Advice:19, chen2019reddit}. 

We also identified a different kind of request focused on {\it technical} support for intended or ongoing IPS, e.g. ``\textit{How can I read my wife's Facebook messages?}''  or ``\textit{How do I use this spyware tool?}'' Many of these users contextualized their asks in a detailed narrative of their situation that included an admission of past IPS, most commonly by reading a partner's text messages or emails. For example, consider this initial post, paraphrased from \deception:

\begin{quoting}[leftmargin = .1 in, vskip= .1 in, indentfirst = false]
    \textit{``I caught my wife by reading her journal and emails. She does not know that I know, and I continue to monitor her email account. I don't think she knows. I haven't told a soul about this, so this is my release. I can elaborate...''}
\end{quoting}

Not all users framed their interactions as requests, however; a subset of thread starters within Forums \marriagebuildersshort\ and \deceptionshort\ posted to share unsolicited advice on working with certain IPS tools. This advice was often couched in a personal narrative (e.g. ``\textit{Here's how I tracked his Internet history after he deleted it}'') and usually promoted the use of the tools. We consider these to be organic advertisements for these tools, the implications of which are discussed in Section~\ref{sec:discussion}.

Overall, we identified three high-level goals for users who sought IPS-related advice across these forums. Many users fixated on reading their partner's emails or text messages, or what we call
\textit{(1) investigation of a partner's prior activities}. Many were also interested in real-time access to their partner's devices, to gain information such as live updates on their partner's location and browsing history: in our view, these attackers sought \textit{(2) continuous monitoring of a partner's current and future device use}. Finally, we saw that many posters expanded the target of their IPS to include a suspected affair partner, with the goal of identifying their personal information (e.g., name, address or vehicle registration). We use the term \textit{affair partner} here to mean the person involved in an affair outside of the intimate relationship, occasionally referred to as the `other' man or woman, and we name this goal as a \textit{(3) compromise of a suspected affair partner's privacy}.

\label{sec:responses}
\paragraph{Community escalations.} Communities' responses to users' requests varied. As expected for online support forums, many responded with emotional support and advice on managing infidelity, including recommendations for looking after users' mental and physical health. In a significant body of threads, however, communities responded by encouraging thread starters to pursue their current enactment of IPS, or even to increase the severity of their attacks. We call this a pattern of \textit{escalation}: a situation in which a user begins with a relatively benign request for information, and through interactions with one or more IPS promoters is presented with ideas for enacting or increasing the severity of an IPS attack.

Consider this example from \deception. A user begins a thread by asking for emotional support: ``\textit{I can't believe my relationship has come to this, but I need some advice. Recently I discovered a situation that I'm not sure how to perceive...}'' An hour later, a responder offers several actionable ways for the user to invade their partner's privacy:

\begin{quoting}[leftmargin = .1 in, vskip= .1 in, indentfirst = false]
    \textit{``There are several things you can do. Start by going into full snoop mode. Purchase a voice activated recorder and put it in his car. Snoop his phone records. Place spyware on his computer. Snoop his emails and FB account.''}
\end{quoting}

From there, a concerning dialogue unfolds. As the thread starter shares more details of their story over follow-up posts, the same responder repeatedly suggests ways to enact IPS, for example by offering: 

\begin{quoting}[leftmargin = .1 in, vskip= .1 in, indentfirst = false]
    \textit{``A voice-activated recorder is cheap, \$40 at Walmart. Stop bringing this up, and make him think everything is back to normal. Then, monitor him. Good luck.''}
\end{quoting}

To get a sense of the prevalence of this thread pattern in our corpus, we conducted an additional qualitative coding effort over our human-labeled sampling of IPS-relevant posts. Three coders first coded a random set of 30 relevant threads (10 from each forum) for whether or not they showed a pattern of escalation. Inter-rater reliability showed substantial agreement between raters: out of the 30 posts, all 3 raters agreed on 25 posts, and the remaining 5 showed 2 out of 3 raters in agreement  (Fleiss's kappa of 0.77). We then split the remaining relevant threads among the three coders, finding that approximately one-third of relevant threads showed patterns of escalation (Table~\ref{tab:dataset}). This proportion remained the same in both forums explicitly focused on investigating suspected deception (\marriagebuildersshort, \deceptionshort), and in more general support forums for those `recovering' from infidelity (\redditshort). 

\paragraph{Community de-escalations.} While escalations appeared with alarming prevalence in our dataset, we also found a handful of instances of the opposite: \textit{de-escalations}, in which the community deterred a user from conducting IPS.

In many of these cases, responders reminded posters of the physical and mental impact of continuously performing IPS on a partner. These responders saw that IPS directly undermined the trust required for recovery of a healthy intimate relationship. 
As this example from \marriagebuilders shows: 
 
\begin{quoting}[leftmargin = .1 in, vskip= .1 in, indentfirst = false]
    \textit{``You've got to ask though, when do you stop snooping? That can't be healthy for your relationship if you're being insecure about everything.''}
\end{quoting}

De-escalating responders also often pointed out that IPS may not help people achieve their goals, and instead sabotage a relationship. For cases where IPS had already been \mbox{committed}, a small number of users pointed out how the intimate partner may be experiencing this level of privacy intrusion. We demonstrate this `pushback' against IPS through this responder on \deception, after a thread starter admits to monitoring his partner through a home security system:

\begin{quoting}[leftmargin = .1 in, vskip= .1 in, indentfirst = false]
    \textit{``You sound crazy to watch her like that! The fact that you've analyzed every little detail on the system tells everyone a lot about your own insecurities ... come on dude, you're trying to make something out of nothing here.''}
\end{quoting}

Some de-escalating responders also reminded thread starters of the potential legal consequences of engaging in IPS. This included warnings that the use of some attacks could result in a criminal record, failed divorce proceedings due to misbehavior, or expulsion from social groups. 

\section{Taxonomy of IPS Attacks}
\label{sec:taxonomy}

We now describe the IPS tools and tactics discussed within these forums. We present a taxonomy (Table~\ref{tab:taxonomy}) of four types of attacks: (1) \textit{tool-based attacks requiring physical access}, including installing spyware on a partner's phone and attaching GPS trackers to their person; (2) \textit{tool-based attacks not requiring physical access}, including leveraging shared cloud accounts; (3) strategies involving \textit{coercion and subterfuge}, for example convincing a partner to provide access, or tricking them into connecting with falsified social media profiles; and (4) \textit{outsourced attacks}, namely hiring private investigators.

\subsection{Tools that require physical access}
\label{sec:taxonomy-tools-physical}

\begin{table}[t]
\centering
\footnotesize
\begin{tabular}{|l|}
\hline
 {\bf Tool-based attacks that require physical access} \\ 
 Using a cellphone backup recovery tool on a partner's device \\
 Installing a keylogger on a partner's device \\
 Installing screen recording spyware on a partner's device \\
 Installing GPS trackers on a partner's body or in their car \\
 Installing audiovisual recorders in the car or the home \\ \hline
 {\bf Tool-based attacks that do not require physical access} \\
 Leveraging features of a shared phone plan \\ 
 Using shared cloud tools to access a partner's personal data \\ 
 Using router monitoring tools to track and manipulate Internet activity \\ 
 Using reverse lookup directories to find personal information \\ 
\hline
 {\bf Coercion and subterfuge} \\ 
 Leveraging physical proximity to gain access \\
 Convincing a partner to give total access \\
 Catfishing a partner \\ \hline
 {\bf Outsourced attacks} \\ 
 Hiring a private investigator \\
\hline
\end{tabular}
\caption{Taxonomy of IPS attacks promoted on these forums.}
\label{tab:taxonomy}
\end{table}

Our analysis surfaced many attacks requiring access to a target's devices. These attacks are particularly possible in IPS, due to the proximity between intimate partners \cite{freed2018stalker}.

\paragraph{Backup recovery tools.} Recall that a common goal for attackers was the discovery of what a partner said in their texts or emails. To this end, responders promoting IPS often recommended the use of \textit{cellphone backup recovery tools}: both specific software dedicated to reading data from phones or SIM cards, and creative workarounds leveraging built-in iOS or Android features to access that same information. Some of the spyware previously reported \cite{chatterjee2018spyware} works by accessing similar data stores; our data show for the first time how attackers share these products with each other, and how they homebrew their own tools for accessing this information. 

In particular, a substantial number of threads were dedicated to tools that recovered deleted texts from iPhones. Similar tools were available for Android phones, and in older threads we even surfaced evidence of responders helping attackers retrieve texts from Blackberries. While some responders in these threads advocated for the use of specific products, others presented instructions for homebrewed tools they had developed to read messages from a partner's backup files synced to shared iTunes or iCloud storage. Some responders posted code anyone could use to convert such backup files into text files for easy reading, and many also offered one-on-one technical support. 

\paragraph{Keyloggers and screen recorders.} Many attackers were interested in continuous capture of their partners' digital activities, such as websites they visited or passwords they used. For these attackers, responders often recommended installing \textit{keyloggers} and \textit{screen recorders} on a partner's devices. These tools had been surfaced as potential spyware in prior work \cite{chatterjee2018spyware}, but our data highlight they are actively shared as solutions for attackers on these forums. One responder on \reddit\ claimed he had installed keyloggers on all PCs and laptops in his home, describing the benefits of these tools:

\begin{quoting}[leftmargin = .1 in, vskip= .1 in, indentfirst = false]
    ``\textit{Great for capturing passwords \& her true thoughts when messaging (things she backspaced over and didn't send).}''
\end{quoting}
Many responders also recommended screen recorders, such as those built for companies to install on workers' devices---in fact, this use case was often invoked to prove a product's legitimacy. Responders also discussed the benefits and drawbacks of specific products, including whether the paid tiers of some tools were worth purchasing. 

\paragraph{Location tracking and audiovisual recording.} We saw many instances of responders recommending tools for environmental surveillance of a partner's activities, conversations and whereabouts, e.g., \textit{voice-activated recorders} and \textit{GPS tracking devices} placed in key locations like a partner's car. Responders were quick to make recommendations about where to obtain these devices, how much one should expect to pay for them, and best practices for hiding them from targets:

\begin{quoting}[leftmargin = .1 in, vskip= .1 in, indentfirst = false]
    ``\textit{A GPS tracker can fit into a purse without them knowing. I'm positive you can figure out a place to stash one in a car. People track autistic kids and animals with them.}''
\end{quoting}

Surveillance of partners in cars was a recurring theme throughout our data. In addition to providing recommendations on the best places in a car to place a GPS tracker, several threads promoted the use of more sophisticated tools that plug into a car's on-board diagnostics (OBD) system and continuously report the car's location to a remote database, to which an attacker can then subscribe. These tools would be useful, one responder said, because ``\textit{unless a person knows to check the OBD they would never think to look for it.}''

\subsection{Tools that do not require physical access} 
\label{sec:taxonomy-tools-no-physical}

For would-be attackers who were unable to access a partner's phone to install spyware or car to plant a GPS tracker, the responders in our data readily provided tools that did not require physical access to partner or device.

\paragraph{Leveraging shared phone plans.}
Many would-be attackers sought ways to leverage the fact that they shared a phone plan with their intended target. Most seemed to know that a partner's call and SMS histories were accessible on a phone bill; in fact, viewing these was often the first thing an attacker tried, and the use of these records as vectors for abuse has been documented \cite{freed2018stalker}. But the contents of messages are often left off of phone bills; in response, our data show these attackers come to the forums to find other ways to obtain more information from their service providers.

Responders regularly provided tips on how to contact service providers and obtain more detailed records: for example, in one thread on \deception, a responder described how to contact Verizon and set up a monthly spreadsheet dump of all call activity. Phone companies were required to provide these records to account owners, the responder claimed, as a form of consumer protection.

Attackers were also savvy to the many other ways a provider's plan management tools could be used to surveil a partner. Verizon, AT\&T, and T-Mobile were purported to have capabilities ranging from email monitoring to mobile keylogging. Consider the following exchange on \marriagebuilders:

\begin{quoting}[leftmargin = .1 in, vskip= .1 in, indentfirst = false]
    {\bf Attacker:} ``\textit{We are in the process of choosing new cellphones and a new company. Which is the best company to keep tabs, records, etc? We currently have iPhones on AT\&T, and their Family Map did help me prove his affair.}'' \\
    \noindent {\bf Responder:} ``\textit{If you're getting everything you need with AT\&T, I would stay with them. They have immediate online access [to phone records] and their GPS is good.}''
\end{quoting}

In this example, we see responders outline the features of shared plans that make them useful to an attacker: immediate online access to call and text histories, quality GPS for location tracking, and family sharing products that provide easy-to-use interfaces for surveillant capabilities. This last type of tool was especially common in our data, confirming prior work \cite{chatterjee2018spyware, freed2018stalker}. This example also highlights the \mbox{collaborative} nature of how attacks surface in these forums, with a responder echoing and encouraging an attacker towards IPS. Of note, the responder in this example is the third-most-prolific superuser of \marriagebuilders, and many of their posts are similarly IPS-related.

\paragraph{Features of shared cloud services.} Many tools that did not require physical device access took advantage of the built-in features of cloud-based sharing tools. The use of cloud tools for abuse has been reported in prior work \cite{freed2018stalker}; however, our data show for the first time how attackers share these tools with each other as ways to overcome targets' defenses. 

In many threads, attackers seemed aware of the ways iCloud tools in particular could be used to surveil partners who had not provided device access. One thread began:

\begin{quoting}[leftmargin = .1 in, vskip= .1 in, indentfirst = false]
``\textit{What is the best spyware if I can't get their phone, but have their Apple ID and password?}''
\end{quoting}

In this example and many others, our data show attackers are encouraged by the forum to use their partner's Apple ID to view their personal messages and photos from a web browser---no device access necessary. This was commonly invoked as a solution for attackers who sought more detailed information on their partner's texts than records from a service provider contained. Many of these attackers reported they arrived at this method of attack because they had seen a drop-off in their partner's texting activity as reported by their phone bills, and had inferred the partner had moved to iMessage or another messaging service that used data rather than SMS. (Messaging that uses data is not typically itemized on a phone bill.)

Some of these attacks, however, did not even require an attacker to use a partner's login, because their personal data was already syncing to a shared Apple device. For example, an attacker on \reddit described discovering she could view a partner's messages on a family iPad, which was synced to her partner's iCloud account. Our data show attacks of this nature also levied against third-parties, namely the affair partners: in one thread on \deception, an attacker describes realizing her partner's affair partner was using an iPad synced to an iCloud account shared by all three-parties, making her purchasing and Internet history accessible for the attacker to browse.

Attackers were particularly eager to share how iCloud tools could be used for location tracking. One attacker on \marriagebuilders described how to use the Significant Locations feature within iOS to examine a partner's recent location history. In another thread on \marriagebuilders, a user shared an article on Find My Friends and called out its abusive potential:

\begin{quoting}[leftmargin = .1 in, vskip= .1 in, indentfirst = false]
    \textit{``Interesting article about an iPhone app called `Find My Friends', which you may be able to load on your spouse's phone to track their whereabouts.''}
\end{quoting}

Cloud-based tools outside of the Apple ecosystem were also called out for similar purposes. One responder on \reddit shared how Android users could view a ``\textit{timeline}'' of a partner's visited locations via their Google Maps account. Another shared how WhatsApp's phone-to-Web syncing features could be used in concert with one-time physical access to maintain continuous access to a partner's messages. This responder described the initial connection as a ``\textit{one-minute job}'' best done while a partner sleeps, and claimed they were ``\textit{actually shocked at what a privacy flaw this seems to be.}''

Lastly, our forums contained many suggestions for mobile spyware products that leveraged cloud-based access to a target's device, such as tools marketed for use in parental control contexts. Much of the discussion of these products also offered advice on free versus paid tiers, setup and configuration, and even best practices for contacting customer service teams. 

\paragraph{Web traffic trackers on shared networks.} In several forums, we discovered threads in which responders offered advice on how to install \textit{web traffic monitoring tools} on a shared WiFi network. The scope of this attack and the level of detail in which it was described was noticeably more sophisticated than others in our data, or what has to our knowledge been previously reported.

In one thread on \reddit, a person who described themselves as a ``\textit{heartbroken techie}'' with a background in software development started a thread detailing how they used a DNS resolution service to monitor the traffic on their home router. With their tool, the attacker said, they could record every website their partner visited, regardless of whether they deleted their Internet history, in the form of reports issued within 24 hours. The attacker shared the command line scripts and configurations they had used, and even offered to share a GitHub repository where others could retrieve their code. In addition to describing how they used the service to monitor router traffic, they went on to discuss how they used its domain blocking alerts to manipulate their partner: 

\begin{quoting}[leftmargin = .1 in, vskip= .1 in, indentfirst = false]
    \textit{``You can set up a customized message (as I call it, the `oh shit' alert) that will pop up if they try accessing a site that is blocked. It's amazing how much someone will confess if they know you're tech-savvy and you tell them you have a detailed history of their actions (even if you don't.)''}
\end{quoting}

In another case, a responder on \marriagebuilders who claimed to be a computer security professional introduced the forum to the concept of a man-in-the-middle attack and recommended an entry-level tool 
for mounting one. As they described, the tool was able to obtain not just a history of websites visited, but also copies of data sent over the network, e.g. the contents of emails and chats. Most notably, they described the tool as a way to actively manipulate a partner's activity:

\begin{quoting}[leftmargin = .1 in, vskip= .1 in, indentfirst = false]
    \textit{``[You can also] modify the data traffic in real time. This can be used for tactics like replacing phone numbers, names and addresses as they travel over the network. Think about creative ways to change the contents of the websites/emails/chats that they're looking at.''}
\end{quoting}

This last example was sourced from one of the `resource threads' in \marriagebuilders. The responder goes on to offer his services to community members who want help mounting such attacks. We discuss the implications of these types of attacks and the role of technologists providing such support in Section \ref{sec:discussion}.

\paragraph{Reverse lookup directories.} Lastly, our data show would-be attackers seeking and receiving tips for investigating their partners' prior actions via reverse lookup tools, used most commonly to identify people from their phone numbers. Most cases presented as a thread starter finding an unknown number in a partner's texts or call records via other attacks, and then asking the forum for advice on how to discover whether it belonged to an affair partner or an escort service:

\begin{quoting}[leftmargin = .1 in, vskip= .1 in, indentfirst = false]
    \textit{``If anyone knows a really good reverse cell lookup, please let me know. Just found a few unknown numbers on my husband's phone.''}
\end{quoting}

Many solutions offered were simple websites containing databases of people's personal information---one thread even offered tips on how to search Facebook by phone number. But responders in our data also recommended a wide array of commercial products that market themselves as collators of public information on individuals (e.g., WhitePages). Many of these tools offer a free tier enabling lookup of names, addresses, and phone numbers in addition to a paid service for more thorough background checks.
Although these tools are relatively unsophisticated from a technical perspective, they featured in several stories that resulted in an attacker confronting their partner or suspected affair partner at an address or phone number located through these services.

\subsection{Coercion and subterfuge} 
\label{sec:taxonomy-coercion}

In addition to recommending specific tools, 
many responders had advice for coercing or subverting a target into providing access to their data and accounts, most often passwords.

\paragraph{Leveraging physical proximity to gain access.}
Attackers frequently shared how they used their close physical proximity to their targets to overcome common defenses without specific tooling. While many of these tactics had been previously reported from victims' perspectives \cite{freed2018stalker}, we report for the first time attackers jointly developing such coercive strategies in public forums.

In many cases, attackers advised each other to manipulate a partner into `accidentally' revealing a password, as seen in the following example from \marriagebuilders:

\begin{quoting}[leftmargin = .1 in, vskip= .1 in, indentfirst = false]
    ``\textit{Get her to send texts \dots while you are sitting next to her. Then try to make out the password as she types it in.}''
\end{quoting}

These strategies often did not require active manipulation. In some cases, gaining access was as simple as waiting for a partner to fall asleep:

\begin{quoting}[leftmargin = .1 in, vskip= .1 in, indentfirst = false]
    \textit{``My wife would get drunk and pass out. It was simple to just hold the iPhone up to her thumb to unlock it. Took pictures of a lot of conversations so I have a record.''}
\end{quoting}

Some would-be attackers sought help creating opportunities like these. In one thread on \marriagebuilders, an attacker asks:

\begin{quoting}[leftmargin = .1 in, vskip= .1 in, indentfirst = false]
    ``\textit{I have wondered if there is a relaxing drug that will knock her out long enough for me to scan her texts and photos. Any suggestions on how I get that phone?}''
\end{quoting}

Once a partner slept or was otherwise unconscious, attackers and responders offered a range of strategies for exploiting their lowered defenses. These included ways to overcome two-factor authentication schemes---namely, resetting passwords on locked accounts and taking the opportunity to capture codes---as well as ways to plant monitoring tools that could track activity long-term, such as swapping their SIM card into a partner's device to capture their call and text activity.

\paragraph{Creating fake profiles to access their social media.} Coercive attacks did not necessarily need to be physical or direct: we also found evidence of manipulations via social media. Several threads showed attacks using fake social media profiles to overcome the privacy controls a target may have set. While the use of fake social media profiles to directly harass targets has been reported \cite{freed2018stalker}, we found evidence of attackers leveraging fake profiles in a new way: using second- and third-degree connections to access a target's profile.

In one thread on \marriagebuilders, a responder details step-by-step how to fabricate a believable Facebook account, and use it to befriend accounts that are friends with the target. The attacker can thus access parts of the target's social media profiles that have been locked away from the attacker, but have remained unlocked to what they believe are friends-of-friends. The responder describes the access afforded thus:

\begin{quoting}[leftmargin = .1 in, vskip= .1 in, indentfirst = false]
    ``\textit{After your friend request has been accepted, revisit the pages where you couldn't see anything before. You'll be shocked at how much information will suddenly be available, as many MANY people set up security so it's not public, but can be viewed by friends-of-friends. In my neck of the woods there are a lot of local bars that have 1000+ friends and guess what? Every one of those 1000+ friends has now given access to those 1000+ people that allow friends-of-friends to see their info, and I'd guess over half do.}''
\end{quoting}

That responder then cautions attackers to ensure the privacy settings on their real and fake Facebook profiles are set to only show information to the account owner, because, the responder says, ``\textit{I don't believe FB is secure.}''

\paragraph{Convincing a partner to provide total access.} Our data also showed attackers and responders championing a strategy of simply convincing a partner that unfettered access to all devices and accounts should be expected in an \mbox{intimate} \mbox{partnership}. In our context, infidelity forums, this was often raised as a way to facilitate reconciliation after an affair: 

\begin{quoting}[leftmargin = .1 in, vskip= .1 in, indentfirst = false]
    ``\textit{Tell her you need her iCloud password to review something. If she refuses, that's a giant red flag. Then I suggest you say `wife, I love you dearly, but if I don't see what's on that phone, then you are telling me that you're cheating. If you have nothing to hide, then let me see it.'} ''
\end{quoting}

Freed et al. \cite{freed2018stalker} previously reported that abusers often convinced IPV survivors to share their passwords during ``good'' phases of a relationship as a way to establish trust, and subsequently threatened them to continue sharing or face consequences when the relationship turned ``bad''. Our work extends this to show that attackers promote to each other the idea of privacy compromise as currency in abusive relationships, often describing this as key to overcoming the emotional toll of suspicions of infidelity. In fact, many of these threads shared stories in which a partner who was suspected of cheating in the past still shared total access months or years later.

\subsection{Outsourced attacks}
\label{sec:taxonomy-outsourcing}

The final category of IPS attacks surfaced in our data are those in which a responder recommends external resources for investigation or monitoring of an intimate partner.

\paragraph{Private investigators (PIs).} As expected for an infidelity forum context, many attackers and responders within our data referenced hiring PIs to track their partners, with the goal of finding evidence of an affair. Notably, hiring a PI was framed as a legal and ethical way to obtain information, as in this example from a responder on \deception:

\begin{quoting}[leftmargin = .1 in, vskip= .1 in, indentfirst = false]
``\textit{You should hire a PI, who acts within the law to obtain the confirmation you require. But that's all that it will be in the vast majority of cases...merely confirmation.}''
\end{quoting}

Many of the PIs recommended in our data offered services within their specific localities, e.g., ``\textit{He does not do surveillance unless the target originates in [specific U.S. state].}'' These recommendations typically included a phone number to call and a person to ask for---or even, in some cases, a person from the forum to claim as a referral. In addition to general recommendations to seek out PIs and referrals to specific PIs, we also found one thread in which a responder posted a link to the website of a specific national agency, as well as another thread with a directory of `\textit{vetted}' PIs.
These recommendations were often framed as more costly than other attacks, to be used as a last resort. One responder on \marriagebuilders\ remarked PIs were expensive, but at least ``\textit{cheaper than a divorce lawyer}''.

\section{Discussion}
\label{sec:discussion}

We now synthesize takeaways from our findings for anti-IPS efforts. First, our work extends the IPS threat model outlined in this and prior works \cite{freed2018stalker, chatterjee2018spyware}. Security experts seeking to prevent their work from misuse in IPS might consider accounting for this threat model in their technology development practices---in particular, Freed et al. \cite{freed2018stalker} supplies the concept of a \textit{UI-bound adversary} as a consideration for design teams.

We additionally outline broader considerations for security experts, including the use of these online communities as a source of IPS threat intelligence and the development of countermeasures that target the commercial entities behind the spyware industry. Finally, we describe how our work is already impacting interventions, and close by posing a set of open ethical questions for the security community.

\paragraph{Online forums are a rich source of IPS threat intelligence.} Our work highlights how analysis of online communities can provide anti-IPS advocates with valuable intelligence on the motivations and tactics of intimate partner abusers. By observing how attackers interact in these forums and the specific tools they promote for use, we were able to surface new knowledge on IPS strategies that can directly inform interventional efforts. 

Our results confirm findings from prior work that showed the abundance of dual-use and overt spyware apps available for attackers \cite{chatterjee2018spyware}, and highlighting victims' experiences of tech-enabled abuse \cite{freed2018stalker}. But the details of how, precisely, abusers learn to mount these attacks had not previously been reported, and attackers' levels of sophistication had not been well-understood. Our analysis highlights how attackers are collaborating on new tactics in these forums, and surfaces how these attacks are conducted at an unprecedented level of granularity. For example, we find that attackers are not just inspecting targets' call histories on shared family plans, as has been reported previously \cite{freed2018stalker}---they are also sharing strategies on how best to contact service providers and obtain more detailed records. We also surface novel attacks more sophisticated than those previously reported in the IPS context, for example the use of WiFi router tools to monitor and manipulate a partner.

Mining these forums for threat intelligence might help anti-IPS efforts stay ahead of attackers' ever-evolving techniques. We see substantial future work in creating semi-automated tools that enable analyses like ours to scale. As our initial keyword searches (Section~\ref{sec:data-desc}) showed, the way IPS manifests in user-generated natural language may be too nuanced for current automated techniques alone to reliably detect. However, human ratings are laborious and inefficient and, importantly, repeated exposure to stories of abuse can inflict harm on people analyzing large bodies of such texts \cite{figley_compassion_1995}. These problems are exacerbated on large social media platforms, where the volume and speed of conversations generated by millions of users creates urgent problems of scale.

We see a role for advanced language processing techniques in overcoming these challenges. For example, a system might quickly and reliably extract the specific strategies recommended within a post without relying on forum-specific features (as applied in cybercrime marketplaces in \cite{durrett2017identifying}). Such a system might provide a valuable pipeline for security and privacy researchers and anti-IPV advocates building frontline defenses against emergent attack strategies, for example the Coalition Against Stalkerware.\footnote{https://stopstalkerware.org}

In fact, our work has already had impact in this regard. We shared our results with the team of practitioners that runs a technology clinic providing direct interventions to IPV survivors facing IPS \cite{havron2019clinical, freed2019clinical}.\footnote{https://www.ipvtechresearch.org/} At time of writing, they are working to integrate our threat intelligence into their training materials for advocates, as well as their clinic's procedures for discovering and mitigating how abusers are enacting surveillance against their clients.

\paragraph{The for-profit industry behind IPS products uses online forums to market their tools.} Our work also shows that the online ecosystems promoting spyware feature a significant presence from companies creating and marketing their own surveillance products. At one extreme, we found entire forums bloated with spam advertisements for a single spyware tool, suggesting these forums were leveraged to manipulate that tool's SEO. We also found that recommendations for IPS tools are not just manifesting as spam advertisements and SEO for specific spyware apps, but are also shared organically among users in the forum communities.

Within these organic posts, responders engage meaningfully with forum-goers' relationship problems, but also evangelize specific IPS tools or approaches and even serve as technical support for users' spyware installations. Their posts reveal the concerns of consumers in this market: we find posts on the merits and drawbacks of a range of spyware products, both free and paid, and the market rates for voice-activated recorders, GPS trackers, and PIs. All told, our analysis suggests these forums are likely one corner of a broad industry offering `solutions' for would-be attackers to turn suspicions of infidelity into actualized IPS. 

These findings suggest a role for mitigation strategies that directly target these commercial entities. Prior work has demonstrated ways to undermine commercially motivated spam and SEO attacks by working directly with banks and payment processors to make e-crime difficult to monetize \cite{kanich2011show, mccoy2012priceless}. Similar approaches may be effective in the context of IPS. Future work should investigate further the mechanics of how forums like these are leveraged as marketing tactics by spyware companies, with the goal of informing such `follow-the-money' countermeasures. Similar countermeasures might be also useful for security experts concerned about keeping their tools from being misused for IPS: it is possible, for example, that dual-use apps \cite{chatterjee2018spyware} are being advertised as spyware. Where possible, security experts should  prevent their tools from being marketed in this way.

\paragraph{Platform-level defenses might mitigate the spread of IPS.} In addition, our work raises concerns for large social media platforms, where conversations escalating into IPS ``how-tos'' may be happening in spaces that are not specifically dedicated to infidelity, surveillance, or intimate relationships.

We have highlighted features of IPS-relevant conversations, and outlined an agenda for creating semi-automated techniques to extract attackers' strategies from such forums. Platforms concerned about their role in enabling the spread of IPS could use this work to develop community norms or content moderation strategies attuned to these forums' dynamics---for example, the moderators of \reddit might consider banning posts that escalate threads into IPS, and instead seek to encourage de-escalation. Future work might investigate further the scale of the problem within popular social media networks as a first step to developing such mitigation strategies.

Social media platforms might also consider the fake profile attacks discussed in Section~\ref{sec:taxonomy-coercion}, and use the patterns we uncovered to more effectively surface falsified accounts. They might also consider de-emphasizing second- and third-degree network connections in users' experiences of their platforms, or offer privacy controls that limit users' audiences to first-degree connections by default.

\paragraph{Online communities are collaboratively creating new IPS attacks.} Our work also shows that people with significant training in computer security and privacy, while potentially well-meaning, are actually helping to further develop IPS attacks on these forums. 

We were surprised to see the level of technical sophistication in some threads, particularly in contrast to the relatively unsophisticated techniques reported in prior studies with victims \cite{matthews2017stories,freed2018stalker}. Some attacks did fall into the bucket of previously known techniques, for example physical privacy violations like shoulder surfing that are addressable through existing defense strategies \cite{khan2018evaluating, davin2017baseline}. However, several involved the use of custom shell scripts and other more sophisticated techniques, including methods for extracting information from artifacts like iPhone backups and leveraging DNS resolution tools to manipulate a partner's Internet traffic.

In a sense, the IPS promoters who championed their homegrown surveillance tools in these forums were engaged in a process of collaborative innovation, working with other tech-savvy community members to create and refine new abuse tactics. This is similar to behaviors seen previously in cybercrime forums~\cite{cybercrime:weis15}, in which a handful of sophisticated users create tools and then provide or sell them to the community. These collaborative processes are not just creating more efficient attacks, they are also making attacks more accessible to more would-be attackers: much of the discussion on these forums serves as how-to guides for less tech-savvy members, and in many cases communities even provide one-on-one troubleshooting. Future research might further analyze the cooperative dynamics of how these forums develop new attacks, and compare these against known tactics used to perpetrate harm in offline settings (c.f., \cite{hearn_violences_1998}).

\paragraph{Infidelity is used as a justification for IPS.} The forums we studied were rife with emotionally vulnerable people seeking and receiving assistance with difficult interpersonal problems. But they were also rife with attackers freely admitting to and promoting the use of surveillance tools against an intimate partner, often by arguing that infidelity justifies surveillance. In this, we see that the context of infidelity both attracted people to the forums as a site for emotional support and masked them from the social exclusion they might have faced if admitting to IPS in a non-infidelity context \cite{hearn_violences_1998}. This is particularly concerning for anti-IPS efforts, as it can set a precedent of using infidelity as an excuse for abusive actions---a practice mirrored in offline discussions with abusers in IPV \cite{nemeth2012sexual}.

We see compelling areas for future work in using these forums to identify the cultural norms and justifications that encourage abusive behaviors. In concert with ongoing behavior change work with abusers \cite{kelly_naming_2016}, such work could draw on the patterns of de-escalation we uncovered to develop alternative strategies for resolving suspicions of infidelity without resorting to IPS. These alternatives could be promoted before IPS on these forums, thereby retaining the supportive community structure they provide to some forum-goers while discouraging abusive practices.

The infidelity forum setting also creates gray areas for the computer security community at large. Some of the responders providing (ab)users with strategies for IPS were self-described computer security experts who reported using the same tools they promoted for surveillance in their professional work. In the context of helping people who reported they were in toxic relationships, these experts may have felt their material support facilitating IPS was justified. What's more, in the context of publicly available forums like the ones in our study, these experts' bespoke surveillance solutions constitute a persistent record accessible to anyone on the Internet with little effort. It is possible that well-meaning computer security experts may have facilitated IPS not just for the user who posted in the forum, but also for the numerous people who would browse these threads in the future.

We raise these tensions as a set of open ethical questions for the security community. Much as medical doctors operate by a professional code of conduct to `do no harm,' should computer security experts abide by a corresponding professional ethos to wield their expertise only for good? How should judgment calls between justifiable and unjustifiable surveillance be made, and who should make them? And how can computer security experts balance publicizing attacks to support anti-IPS efforts (for example, in this work) against the possibility that doing so might inadvertently help more attackers?

\section{Conclusion}

We have provided the first measurement study of the online communities in which people enacting IPS discuss their tactics. Through a mixed-methods study of five public forums, including a Reddit subforum dedicated to infidelity, we developed a taxonomy of the IPS strategies attackers discuss online. We showed that these forums are sites for both spam advertising specific spyware products and organic discussion of surveillance tools between users. We highlighted threads in which (ab)users learn new IPS tactics from more tech-savvy forum-goers, as well as cases where forums deter them from conducting IPS. Our work is already impacting anti-IPS interventions by informing programs that directly assist victims.

\section*{Acknowledgments}

We thank Sandra Ebirim for vital contributions to the data analysis phase of our study. We are also grateful to our reviewers, whose comments greatly helped to improve our paper. This work was funded in part by NSF Awards \#1916096 and \#1916126, as well as gifts from Facebook and Google.

\bibliographystyle{plain}
\bibliography{references.bib}

\appendix
\newpage
\section{Codebook}
\label{sec:codebook}

Our thematic analysis (see Section \ref{sec:qual-methods}) resulted in the following codebook:

\begin{table}[ht]
\centering
\small
\begin{tabular}{|l|l|}
\hline
\multicolumn{1}{|l|}{\textbf{Category}} & \textbf{Code} \\ \hline
\multirow{12}{*}{forum culture} & "hacking" \\
 & do not use this hacker / tool \\
 & emotional advice \\
 & encouraging transparency \\
 & giving permission for IPS \\
 & moderation \\
 & norm enforcement \\
 & referring to another thread \\
 & technical advice \\
 & using slang \\
 & escalation \\
 & de-escalation \\ \hline
\multirow{16}{*}{tools and tactics} & accessing accounts / getting around passwords \\
 & accessing backups \\
 & accessing phone records / cellphone bills \\
 & accessing physical device \\
 & accidental surveillance \\
 & audiovisual recording / listening / watching \\
 & cloud / shared accounts \\
 & ways to get around 2FA / victims being notified \\
 & hacker / tool ad \\
 & location tracking \\
 & paid tools \\
 & reading emails / texts \\
 & remote attacks \\
 & social media accounts \\
 & search Google with these keywords \\
 & tracking browsing activity \\
 & tracking keystrokes / keylogger \\ \hline
\end{tabular}
\caption{Codebook resulting from thematic analysis of IPS-relevant forum postings (see Section \ref{sec:qual-methods}).}
\label{tab:codebook}
\end{table}

\end{document}